\documentclass{article}

\usepackage{geometry}
\usepackage{amsbsy,amsfonts,amsmath,amssymb,enumerate,epsfig,graphicx,rotating,subfigure}
\usepackage{makeidx}
\usepackage{multicol}
\usepackage{multirow}
\usepackage{url}
\usepackage{algorithm}
\usepackage{algorithmicx}
\usepackage[noend]{algpseudocode}
\usepackage{listings}
\usepackage{cleveref}
\usepackage{wrapfig}
\usepackage{tikz}
\usepackage{standalone}
\usepackage{bm}
\usepackage{bbm}
\usepackage[font={footnotesize,it}]{caption}
\usepackage{makecell}
\usepackage{upgreek}
\usepackage{siunitx}
\usepackage{autonum}
\usepackage{setspace}
\usepackage{enumitem}

\usetikzlibrary{calc}
\usetikzlibrary{decorations.markings}
\usetikzlibrary{shapes,snakes}
\usetikzlibrary{shapes.geometric}

\usepackage[acronym]{glossaries}
\makeglossaries

\newacronym{hhg}{HHG}{hierarchical hybrid grids}
\newacronym{tme}{TME}{textbook multigrid efficiency}
\newacronym[plural=PDEs]{pde}{PDE}{partial differential equation}
\newacronym{pspg}{PSPG}{Pressure Stabilized Petrov-Galerkin}
\newacronym{fmg}{FMG}{full multigrid}
\newacronym{flops}{FLOPS}{floating point operations per second}
\newacronym{wu}{WU}{work unit}
\newacronym{mumps}{MUMPS}{MUltifrontal Massively Parallel sparse direct Solver}
\newacronym{petsc}{PETSc}{Portable, Extensible Toolkit for Scientific Computation}
\newacronym{ecm}{ECM}{Execution-Cache-Memory}

\usepackage[utf8]{inputenc} 
\usepackage{xspace}

\newcommand{\hyteg}{\mbox{\textsc{HyTeG}}\xspace}
\newcommand{\likwid}{\mbox{\textsl{likwid}}\xspace}

\newcommand{\supermucng}{\mbox{SuperMUC-NG}\xspace}

\newcolumntype{H}{>{\setbox0=\hbox\bgroup}c<{\egroup}@{}}

\newcommand {\gradOp} {\nabla } 
\newcommand {\divOp} {\nabla \cdotp} 

\newcommand\restr[2]{{
  \left.\kern-\nulldelimiterspace 
  #1 
  \vphantom{\big|} 
  \right|_{#2} 
}}

\newcommand{\norm}[1]{||#1||}

\newcommand{\diag}{\mathop{\mathrm{diag}}}

\newcommand{\eg}{\mbox{e.\,g.}\xspace}
\newcommand{\ie}{\mbox{i.\,e.}\xspace}

\newcommand{\cf}{\mbox{cf.}\xspace}

\begin{document}

\title{Textbook efficiency: massively parallel matrix-free multigrid for the Stokes system}
\author{%
Nils Kohl\thanks{Chair for System Simulation (LSS), Friedrich-Alexander Universit\"at Erlangen-N\"urnberg, Germany (\protect\url{nils.kohl@fau.de}, \protect\url{ulrich.ruede@fau.de})} 
\and%
Ulrich R\"ude\footnotemark[1] \thanks{Centre Europ\'een de Recherche et de Formation Avanc\'ee en Calcul Scientifique (CERFACS), France}%
}

\date{}
\maketitle

\begin{abstract} 
We employ \gls*{tme}, as introduced by Achi Brandt,
to construct an asymptotically optimal monolithic multigrid solver for the Stokes
system. 
The geometric multigrid solver builds upon the concept of \gls*{hhg}, which is
extended to higher-order finite-element discretizations, and a corresponding
matrix-free implementation. 
The computational cost of the \gls*{fmg} iteration 
is quantified, and the solver is applied to multiple benchmark problems.
Through a parameter study, we suggest configurations that achieve \gls*{tme} for both,
stabilized equal-order, and Taylor-Hood discretizations. 
The excellent node-level performance of the relevant compute kernels
is presented via a roofline analysis. Finally, we demonstrate the weak and strong 
scalability to up to $147,456$ parallel processes and solve Stokes systems with 
more than $3.6 \times 10^{12}$ (trillion) unknowns.
\end{abstract}



\glsresetall

\section{Introduction}\label{sec:introduction}

\Gls*{tme}, a term coined by Achi Brandt in \cite{brandt1998barriers,thomas2003textbook}, 
suggests that
an ideal multigrid algorithm 
should solve a discrete system with less than $10$ times the computational
work that is required to apply the corresponding operator.

The computational work $\mathfrak{W}(M)$ required to employ a numerical method $M$, in order to
solve a linear system $Ax=b$, is conveniently
expressed in multiples of a \gls*{wu}. 
One \gls*{wu} amounts to the computational work $\mathfrak{W}(A)$ 
required for application of the considered linear operator, \ie
\begin{align}\label{eq:work-unit}
    1\text{WU} := \mathfrak{W}(A).
\end{align} 
Consequently, we achieve \gls*{tme} if we design a 
multigrid method $\text{MG}$, that solves $Ax = b$, with
\begin{align}\label{eq:tme}
    \frac{\mathfrak{W}(\text{MG})}{\mathfrak{W}(A)} < 10.
\end{align}

We emphasize that \gls*{tme} is defined
with respect to the underlying \emph{differential} equation.
Solving the \gls*{pde} with optimal complexity is a more ambitious goal 
and is potentially more difficult to achieve than 
only showing the mesh independent \emph{algebraic} convergence of an iterative solver.
To evaluate whether an algorithm reaches \gls*{tme},
also the specific \gls*{pde} must be well defined, including the (class of) boundary conditions and forcing terms 
under consideration.

With a proclaimed cost of only 10 \gls*{wu}, the \gls*{tme} paradigm
sets a concrete quantitative limit of the cost for solving the \gls*{pde}.
In this sense, reaching \gls*{tme} is again more challenging
than the notion of asymptotic optimality as it is typically shown in
abstract multigrid convergence results. 
In the latter case the constants remain unspecified so that no comparison 
of the efficiency between different methods is possible.
However, though fundamentally desirable, the strong notion of
\gls*{tme} efficiency is yet rarely discussed in the literature
and, in fact, has not been demonstrated for many algorithms or problems.

For this article, we specifically point out that to reach \gls*{tme} for higher order discretizations
must be expected to be more difficult than for low order discretizations.
We identify three effects that will usually drive up the cost for solving linear systems in a \gls*{tme} setting
for higher order discretizations.
\begin{itemize}[topsep=4pt, itemsep=4pt, partopsep=4pt, parsep=4pt]
\item
    First, we must not forget that the cost of a \gls*{wu} itself will typically increase with
    the order of discretization, most typically because a plain higher order 
     stiffness matrix will have more nonzero entries and is thus more densely
    populated. 
\item
    The linear system for higher order discretization may be less well suited for an iterative solution,
    typically it exhibits a higher condition number so that, e.g., Krylov methods may need more iterations to 
    reach the same error tolerance as for equivalent low order systems
\item
    Finally, and possibly most intriguingly, the expected cost for a nested iteration will grow.
    While for a typical $\mathcal{O}(h^2)$ discretization, the error is expected to improve asymptotically by a factor of $4$ per
    level of canonical mesh refinement $h \rightarrow h/2$, this factor increases
    to $16$ for a $\mathcal{O}(h^4)$ discretization.
    This means that in \gls*{fmg} \cite{brandt2011multigrid}, each new level for a high order discretization 
    must enforce also much stricter algebraic convergence than for low order.
    Where a single v-cycle on each level may be enough for low order, for high order it may
    well be that two or more v-cycles will become necessary.
\end{itemize}

Additionally, we note that the denser matrix structure and higher flop cost of higher order
discretizations is sometimes considered to be an advantage, since
it leads to more flops per memory operations in the iterative solver.
This is then termed a better \emph{algorithmic balance} \cite{Hager2010introduction}.
A better algorithmic balance may help to better exploit computer architectures
whose memory throughput  is poor compared to their peak \gls*{flops}.

The solvers considered in this article are built on top of the \gls*{hhg} structure 
\cite{bergen2004hierarchical,Bergen2006Diss,gmeiner2015performance,
bauer2018,gmeiner2016quantitative,gmeiner2015towards,bauer2020terraneo}.
In previous work, the efficiency, performance, and extreme scalability of matrix-free multigrid 
solvers based on \gls*{hhg} have been demonstrated for linear, nodal discretizations.
A main contribution of this paper is, besides the construction of an efficient geometric 
multigrid solver, the extension of \gls*{hhg} to higher-order discretizations.

The SIAM review article \cite{ruede2018research}
provides a broad overview over the relevance of fast numerical methods for the solution of \glspl*{pde}.
Besides the standard literature on multigrid solvers 
\cite{brandt2011multigrid,trottenberg2000multigrid,hackbusch1994iterative}, there is exhaustive
research that cannot be referenced in its entirety. Various smoothers for the solution of elliptic 
problems are presented in \cite{vanka1986block,braess1997efficient,adams2003parallel,may2015scalable,verfurth1984multilevel}.
The relevance of matrix-free methods is underlined \eg in 
\cite{bastian2019matrix,brown2010efficient,kronbichler2012generic,may2015scalable,rudi2015extreme}.
Domain specific languages and code generation are rather current approaches towards fast iterative solvers \cite{kuckuk2016automatic}.
Apart from geometric multigrid solvers, algebraic multigrid methods are an efficient alternative, especially for
problems on unstructured grids \cite{baker2016scalability}, as well as massively solvers based on domain decomposition
\cite{chan1994domain,klawonn2015toward,wohlmuth2001iterative}.

In this article, we consider the Stokes system as model problem. Discretization with
finite-elements yields a linear system with saddle-point structure. \Cref{sec:discretization}
covers the discretization and introduces the parallel data structures of the matrix-free implementation.
After definition of the multigrid components in \cref{sec:multigrid}, we quantify the computational cost
of the resulting \gls*{fmg} iteration in \cref{sec:tme}. Finally, in \cref{sec:performance}, we present
numerical benchmarks to find optimal solver configurations in the sense of \gls*{tme}, and demonstrate
the node-level performance as well as the parallel scalability of the implementation.

\section{Finite element discretization for the Stokes system}\label{sec:discretization}

As model problem we consider the constant-coefficient Stokes system that describes viscous fluid 
motion on a bounded, polyhedral domain $\Omega \subset \mathbb{R}^3$, defined by
\begin{align}\label{eq:stokes-cc}
- \Delta \mathbf{u} + \gradOp p = \mathbf{f} \\
\divOp \mathbf{u} = 0
\end{align}
where $\mathbf{u} = (u_1, u_2, u_3)^\top$ represents the vector-valued velocity field, 
$p$ the scalar pressure field, and $\mathbf{f} = (f_1, f_2, f_3)^\top$ an external
force acting on the fluid. 
We consider Dirichlet, and natural Neumann outflow boundary conditions on 
$\partial \Omega = \partial \Omega_D \cup \partial \Omega_N$
\begin{align}
    \mathbf{u} = \mathbf{w} \ \text{on} \ \partial \Omega_D, \quad %
    \frac{\partial \mathbf{u}}{\partial \mathbf{n}} = \mathbf{n} p \ \text{on} \ \partial \Omega_N
\end{align}
where $\mathbf{n}$ is the outward pointing normal at the boundary.
If $\partial \Omega = \partial \Omega_D$, the pressure is defined up to a constant and the
Dirichlet boundary function $\mathbf{w}$ must satisfy compatibility conditions \cite{elman2014finite}. 
We fix $p$ to a mean value of $0$ by setting $\int_\Omega p\, \mathrm{d}x = 0$. 

\begin{figure}[h]
    \centering
    \includegraphics[width=0.8\textwidth]{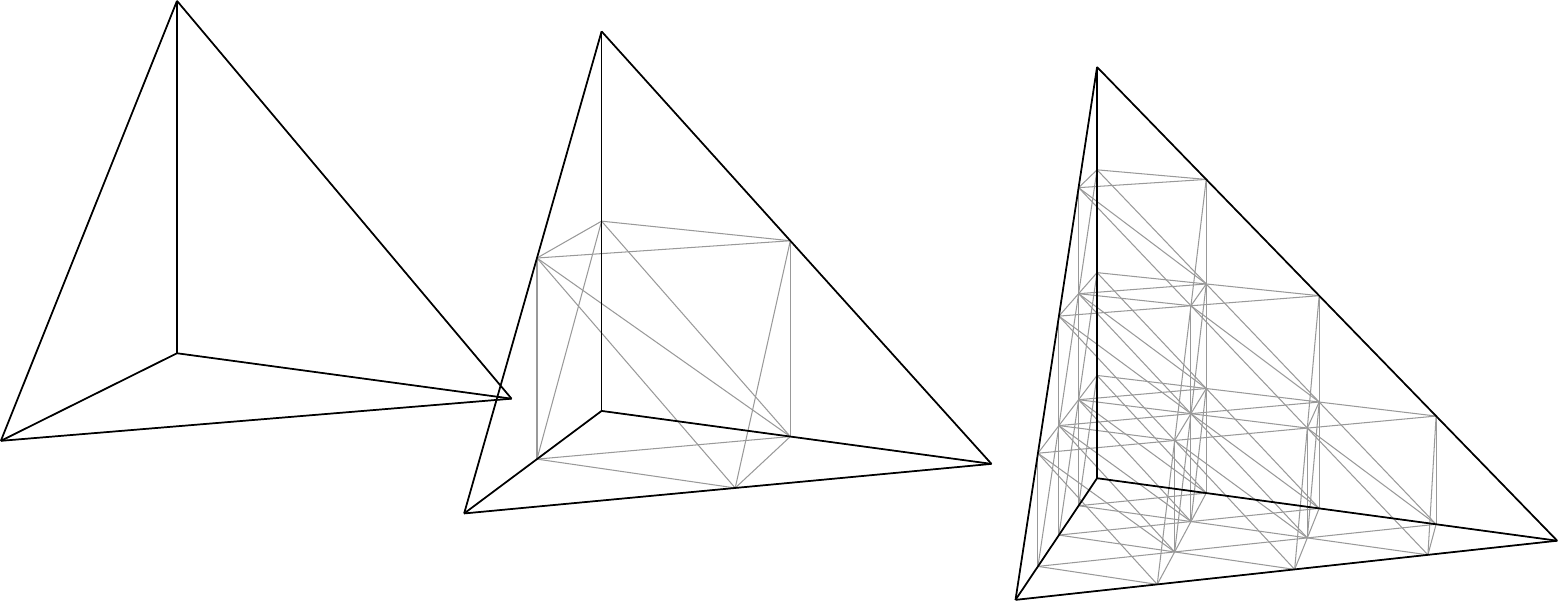}
    \caption{Uniform, structured refinement of a single tetrahedron of
    the unstructured mesh. From left to right: initial tetrahedron, 
    refinement level $\ell = 1$, refinement level $\ell = 2$.}
    \label{fig:hhg-refinement}
\end{figure}

Let $\mathcal{T}_0$ denote an unstructured partitioning of the computational domain into tetrahedral elements.
Each of the elements in $\mathcal{T}_0$ is then successively and uniformly refined according to 
\cite{bey1995tetrahedral}, yielding a hierarchy of tetrahedral meshes
$\mathcal{T} = \{\mathcal{T}_\ell, \ell = 0, ..., L\}$.
The structured refinement of a single coarse grid tetrahedron 
is illustrated in \cref{fig:hhg-refinement}.
The mesh hierarchy is discussed in more detail in \cref{sec:hhg,sec:edgedofs}.

We define the solution and test spaces $\mathbf{H}^1_E$ and $\mathbf{H}^1_{E_0}$
\begin{align}
    \mathbf{H}^1_E := \{ \mathbf{u} \in \mathcal{H}^1(\Omega)^3 : \mathbf{u} = \mathbf{w} \ \text{on} \ \partial \Omega_D \}, \quad 
    \mathbf{H}^1_{E_0} := \{ \mathbf{v} \in \mathcal{H}^1(\Omega)^3 : \mathbf{v} = 0 \ \text{on} \ \partial \Omega_D \},
\end{align}
and standard conforming finite element spaces $\mathbf{V}^\ell_0 \subset \mathbf{H}^1_{E_0}$, 
$\mathbf{V}^\ell_E \subset \mathbf{H}^1_{E}$, and $Q^\ell \subset L_2(\Omega)$ defined by polynomial 
functions on each tetrahedron for each level $\ell$ of the mesh hierarchy.

The standard weak formulation of \cref{eq:stokes-cc} is \cite{elman2014finite}: 
find $\mathbf{u}_\ell \in \mathbf{V}^\ell_E$ and $p_\ell \in Q^\ell$ with
\begin{equation}\label{eq:stokes-weak}
\begin{aligned}
    \int_\Omega \nabla \mathbf{u}_\ell : \nabla \mathbf{v}_\ell - \int_\Omega p_\ell \ \nabla \cdot \mathbf{v}_\ell 
    &= \int_\Omega \mathbf{f} \cdot \mathbf{v}_\ell \quad \text{for all } \mathbf{v}_\ell \in \mathbf{V}^\ell_0 \\
    \int_\Omega q_\ell \ \nabla \cdot \mathbf{u}_\ell &= 0 \quad \text{for all } q_\ell \in Q^\ell.
\end{aligned}
\end{equation}

With basis functions $\{ \boldsymbol{\upphi}_j \}$ of $\mathbf{V}^\ell_E$ and $\{ \psi_j \}$ of $Q^\ell$, we define
coefficient vectors $\underline{\mathbf{u}}_\ell \in \mathbb{R}^{n_u}$, $\underline{\mathbf{p}}_\ell \in \mathbb{R}^{n_p}$, 
$\underline{\mathbf{f}}_\ell \in \mathbb{R}^{n_u}$, and $\underline{\mathbf{g}}_\ell \in \mathbb{R}^{n_p}$ by
\begin{align}
 \mathbf{u}_\ell = \sum_{j=1}^{n_u + n_\partial} \underline{\mathbf{u}}_{\ell, j} \boldsymbol{\upphi}_j, \quad 
 p_\ell          = \sum_{j=1}^{n_p} \underline{\mathbf{p}}_{\ell, j} \psi_j,\\
\end{align}
where $\sum_{j=1}^{n_u} \underline{\mathbf{u}}_{\ell, j} \boldsymbol{\upphi}_j \in \mathbf{V}^\ell_0$, and
$\sum_{j=n_u + 1}^{n_\partial} \underline{\mathbf{u}}_{\ell, j} \boldsymbol{\upphi}_j$ interpolates the boundary function $\mathbf{w}$
on $\partial \Omega_D$.

The discrete formulation of \cref{eq:stokes-weak} on level $\ell$
results in the saddle point problem
\begin{align}\label{eq:stokes-discrete}
\mathcal{A}_\ell
\begin{bmatrix}
\underline{\mathbf{u}}_\ell \\
\underline{\mathbf{p}}_\ell
\end{bmatrix}
=
\begin{bmatrix}
\underline{\mathbf{f}}_\ell \\
\underline{\mathbf{g}}_\ell
\end{bmatrix},\quad
\mathcal{A}_\ell=
\begin{bmatrix}
\mathbf{A}_\ell & \mathbf{B}_\ell^\top \\
\mathbf{B}_\ell & -C_\ell
\end{bmatrix}, \quad
\mathbf{A}_\ell = \diag(A_{x, \ell}, A_{y, \ell}, A_{z, \ell})
\end{align}
where $\mathbf{A}_\ell \in \mathbb{R}^{n_u \times n_u}$ is the discrete, block-diagonal vector-Laplacian,
$A_\ell := A_{x, \ell} = A_{y, \ell} = A_{z, \ell}$ the scalar Laplacian, and $\mathbf{B}_\ell \in 
\mathbb{R}^{n_p \times n_u}$ represents the divergence matrix, with
\begin{align}
    \mathbf{a}_{ij} = \int_\Omega \nabla \boldsymbol{\upphi}_i : \nabla \boldsymbol{\upphi}_j, \quad 
    \mathbf{b}_{kj} = - \int_\Omega \psi_k \nabla \cdot \boldsymbol{\upphi}_j.
\end{align}
The right hand side of \cref{eq:stokes-discrete} is given by
\begin{align}
    \underline{\mathbf{f}}_{\ell, j} = \int_\Omega \boldsymbol{\upphi}_j \cdot \mathbf{f} -  \sum_{j=n_u + 1}^{n_u + n_\partial} \underline{\mathbf{u}}_{\ell, j} \int_\Omega \nabla \boldsymbol{\upphi}_i : \nabla \boldsymbol{\upphi}_j, \quad 
    \underline{\mathbf{g}}_{\ell, k} = - \sum_{j=n_u + 1}^{n_u +{} n_\partial} \underline{\mathbf{u}}_{\ell, j} - \int_\Omega \psi_k \nabla \cdot \boldsymbol{\upphi}_j. 
\end{align}

The approximation spaces $\mathbf{V}^\ell_0$, $\mathbf{V}^\ell_E$, and $Q^\ell$ must be chosen carefully, to assert
uniform inf-sup stability \cite{elman2014finite}. Two common choices are stabilized equal-order, linear finite elements
for velocity and pressure ($\mathbf{P}_1-\mathbf{P}_1$ approximation) \cite{hughes1986new}, and the stable pairing of quadratic elements
for the velocity and linear elements for the pressure ($\mathbf{P}_2-\mathbf{P}_1$ approximation), also referred to
as \emph{Taylor-Hood} method \cite{taylor1973numerical}.
For the stable $\mathbf{P}_2-\mathbf{P}_1$ approximation, $C_\ell = 0$ in \cref{eq:stokes-discrete}.
The $\mathbf{P}_1-\mathbf{P}_1$ approximation requires additional stabilization. We employ the 
\gls*{pspg} stabilization \cite{brezzi1988stabilized,hughes1986new} by setting $C_\ell$ according to
\begin{align}\label{eq:pspg}
    c_{ij} = \sum_{T \in \mathcal{T}_\ell} \delta_T h^2_T \int_T \nabla \psi_i \cdot \nabla \psi_j, \quad i,j = 1, \dots, n_p
\end{align}
where $h_T = (\int_T \mathrm{d}x)^{1/3}$ depends on the size of the element $T$, and $\delta_T$ is chosen as $1/12$ \cite{drzisga2018analysis}.
The right-hand side block vector $\underline{\mathbf{g}}_\ell$ must be modified accordingly at the Dirichlet boundary \cite{drzisga2018analysis}.

\subsection{Hierarchical hybrid grids}\label{sec:hhg}

Support for unstructured grids is arguably one of the major advantages
of the finite element method. 
However, unstructured grids may complicate the implementation of geometric multigrid 
methods, and efficient matrix-free compute kernels 
that are essential 
for solving extreme-scale problems \cite{bauer2018,gmeiner2016quantitative}.
Fully structured meshes, on the other hand, often cannot adequately represent real-world domains.
Hierarchical hybrid grids (HHG) \cite{bergen2006massively,bergen2004hierarchical,Bergen2006Diss,gmeiner2015performance,bauer2020terraneo} 
are one attempt to combine the best of both worlds.

During the refinement of the coarsest grid $\mathcal{T}_0$, all 
new tetrahedra are geometrically
identical (up to translation) to one out of six different tetrahedra \cite{bergen2006massively,bey1995tetrahedral}.
Also, the edges at each inner node of the mesh span a geometrically
identical \emph{stencil} \cite{Bergen2006Diss,gmeiner2015performance}.
The stencil at an unknown is defined by its coupling to other unknowns, 
potentially including itself. 
In the finite element context, this 
corresponds to the non-zero entries of a row in the globally assembled
system matrix.
The invariant topology of the stencil in the uniformly refined
tetrahedron has some important implications. 
In particular, unknowns can be updated in a 
node-centered, stencil-based fashion. 
This may result in performance advantages and it simplifies the realization of
operators that require the diagonal entries of the system matrix (\eg Gauss-Seidel smoothers) \cite{gmeiner2015performance}.
Additionally, for operators with constant coefficients, the stencil values are constant at each node
inside a refined coarse grid tetrahedron, providing even more optimization
opportunities. 
Moreover, the successive refinement yields a grid hierarchy suitable 
for geometric multigrid 
methods by design.

\begin{figure}[ht]
    \centering
    \includegraphics[width=0.8\textwidth]{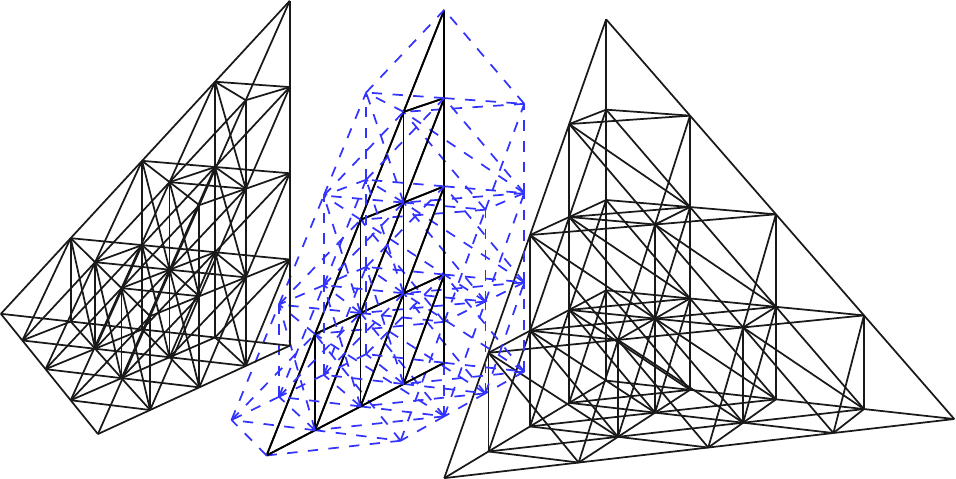}
    \caption{Macro-face primitive that interfaces two macro-cells. Each
    of the three primitives can be distributed to arbitrary parallel processes.
    The dashed lines indicate ghost-layer data.}
    \label{fig:hhg-macro-face}
\end{figure}

The block-structured domain partitioning allows defining entirely distributed data structures
that are crucial for extreme-scalable simulations \cite{bauer2020walberla,godenschwager2013framework}.
To communicate data between the refined tetrahedra (also called macro-cells), 
we introduce \emph{interface primitives} \cite{bergen2006massively,bergen2004hierarchical,kohl2019hyteg}.
For each face, edge, and vertex of the unstructured coarse mesh, a
separate face-, edge-, or vertex-primitive data structure is created.
In the following we will refer to them as  macro-faces, macro-edges and macro-vertices, respectively.
\Cref{fig:hhg-macro-face} shows
a macro-face primitive that interfaces two neighboring macro-cells.
Each macro-primitive, including the interface primitives, is assigned to 
exactly one of the parallel processes.
Note that one process may thus hold multiple 
macro-primitives. 
Each grid point of the refined mesh is assigned to exactly one macro-primitive.
To allow for read access to neighboring unknowns, the macro-primitives
are extended by ghost-layers.
When necessary, i.e., usually before operator applications, the ghost-layers
are updated by communication between neighboring macro-primitives.

The primitives are connected, such that they can be interpreted as a graph, where the
graph-vertices correspond to the primitives, and the graph-edges correspond to the
communication paths, as described in \cite{kohl2019hyteg}.
This way, scalable load balancing algorithms similar
to those discussed in \cite{burstedde2011p4est,schornbaum2018extreme}
can be straightforwardly applied to the graph.
For a more detailed discussion of the implementation of the parallel data structures
we refer to \cite{Bergen2003HHG,bergen2004hierarchical,kohl2019hyteg}.

\subsection{Generalization of HHG towards non-nodal discretizations}\label{sec:edgedofs}

The performance and scalability of the HHG prototype framework
and the corresponding data structures
has been studied in \cite{bergen2006massively,bergen2004hierarchical,Bergen2006Diss,
gmeiner2016quantitative,gmeiner2015performance,gmeiner2015towards};
matrix-free methods and 
applications have been discussed in \cite{bauer2018stencil,bauer2018,bauer2017two} and 
\cite{huber2018surface,waluga2016mass,weismuller2015fast}.
These articles have demonstrated that HHG 
can reach scalability on the 
largest available core and node counts.
However, so far the HHG principle was limited to low order nodal finite elements.

A major contribution of this article is the presentation of extended HHG data structures 
and to demonstrate how the HHG 
paradigm can
be extended and augmented to support more general discretizations.
A crucial point here is that 
the uniform refinement can be mapped to efficient linearized data structures.
This becomes a key step to implement
efficient kernel routines for all data 
that is located on the edges, faces, or within a cell of each grid element.
In this article, we focus on the realization of edge-centered unknowns in the HHG architecture. 
Based on this software extension, we then implement quadratic, conforming finite elements. 
We point out that the described concepts can be
transferred also to face or cell-unknowns for further generalizations and as they 
are needed for other finite-element discretizations.

\begin{figure}[ht]
    \centering
    \subfigure[at vertex]{\label{fig:p2-stencils-pt-1-vertex}
        \resizebox*{0.45\textwidth}{!}{
            \includegraphics{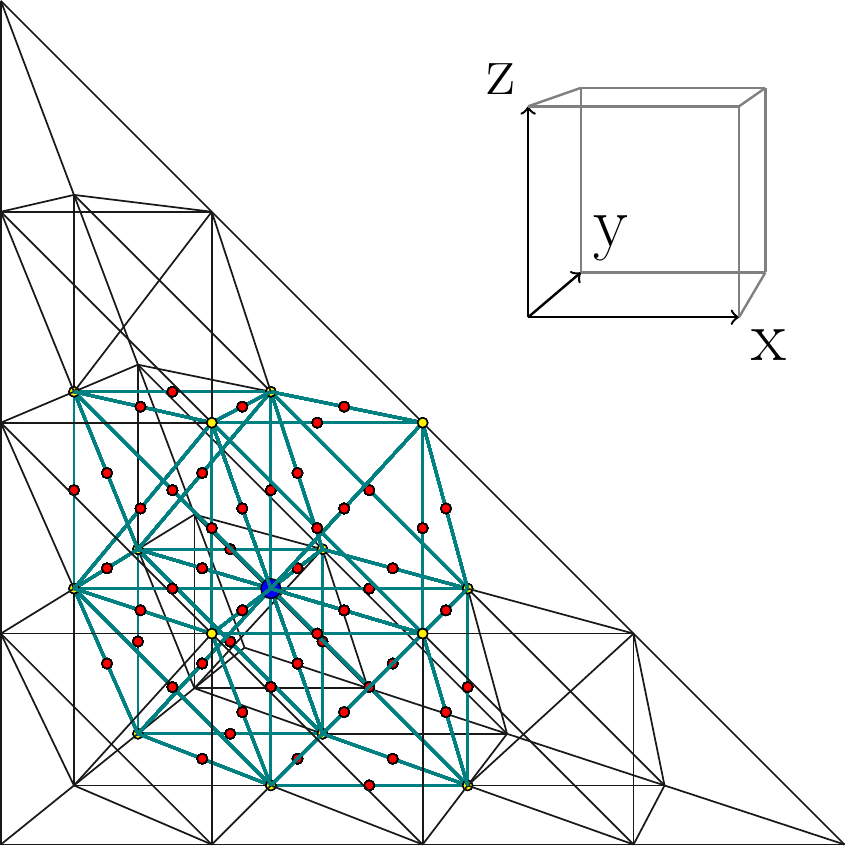}
    }}%
    \hfill
    \subfigure[at x-edge]{
        \resizebox*{0.45\textwidth}{!}{
            \includegraphics{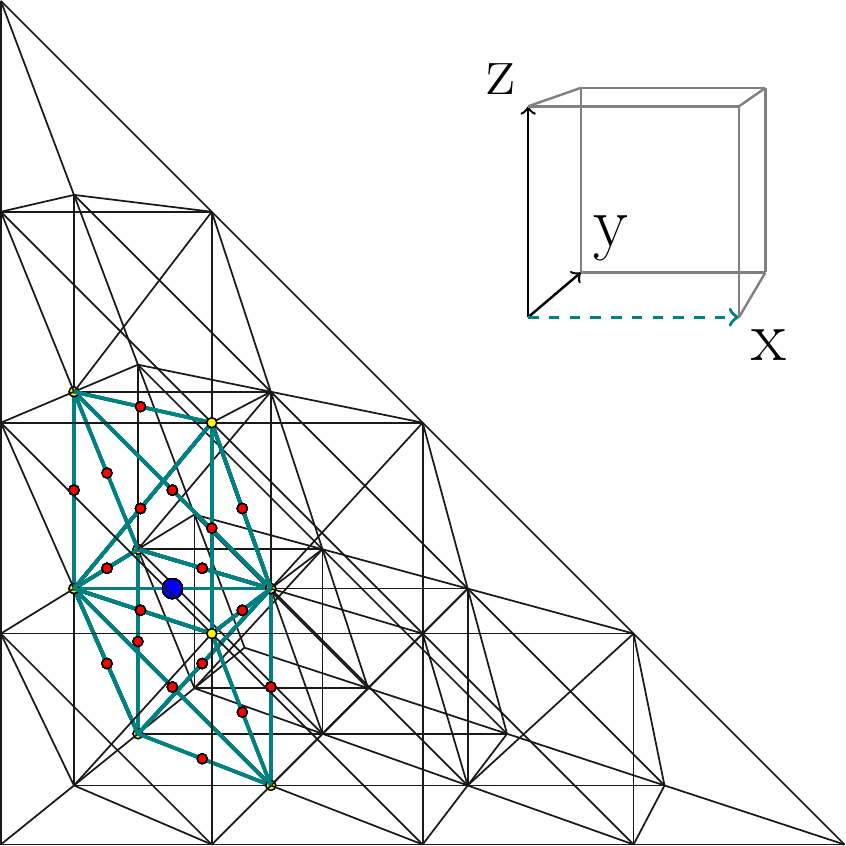}
    }}%

    \subfigure[at yz-edge]{
    \resizebox*{0.45\textwidth}{!}{
        \includegraphics[width=\textwidth]{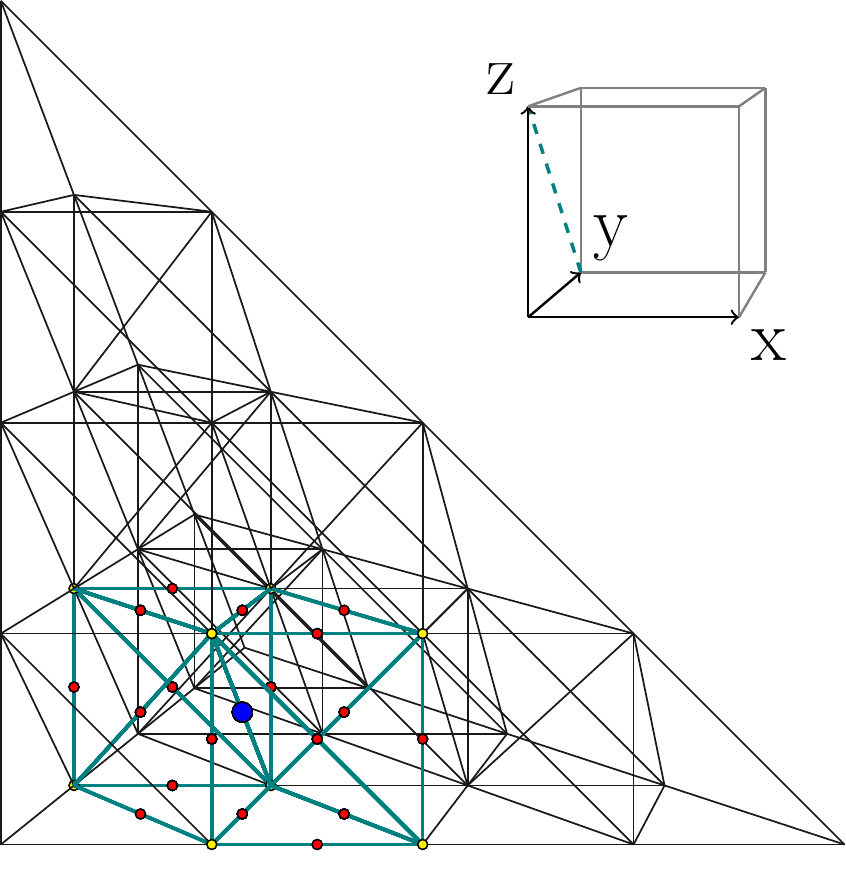}
    }}%
    \hfill
    \subfigure[at xyz-edge]{
        \resizebox*{0.45\textwidth}{!}{
            \includegraphics[width=\textwidth]{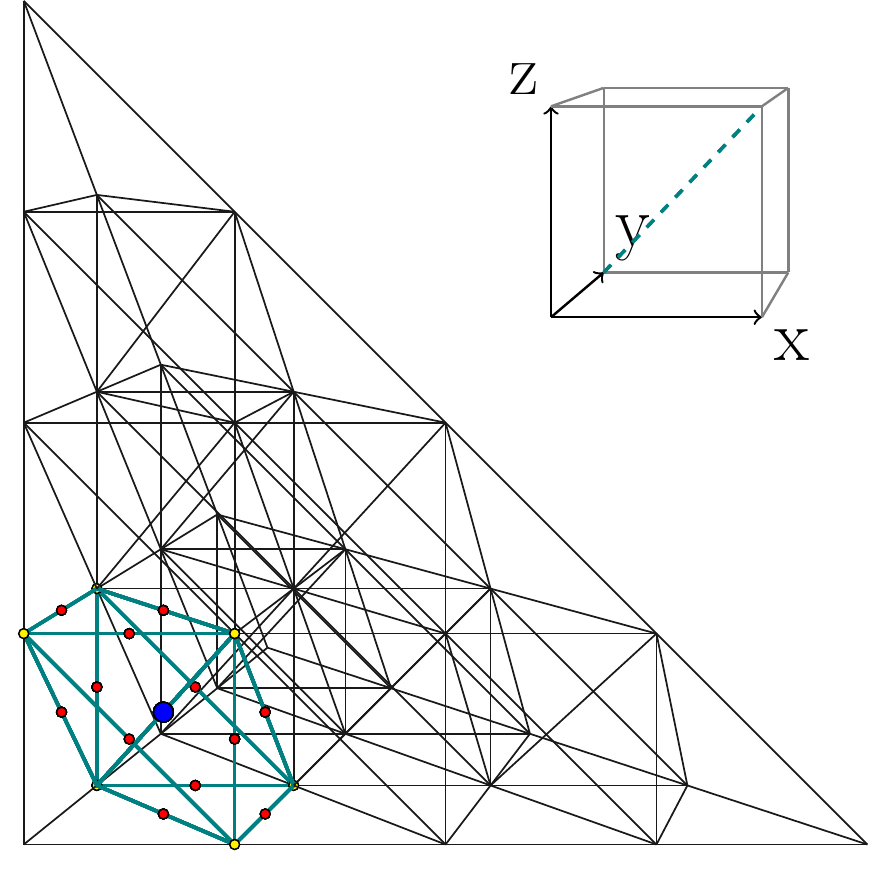}
    }}%
    \caption{Neighboring elements and corresponding stencils at vertex and edge unknowns for 4 of 8 required stencils
    (vertex-, x-, yz-, and xyz-centered).}
    \label{fig:p2-stencils}
\end{figure}

To reveal their uniform structure, we split the edges of a refined tetrahedron into seven subgroups,
corresponding to the seven different orientations of edges that are created during the refinement.
We denote the groups as x-, y-, z-, xy-, xz-, yz-, and xyz-edges - named after the coordinate directions
in which they are spanned.
Each of these subgroups shares the properties of the nodal layout of the HHG meshes:
the neighboring elements of each edge-unknown of one group have identical geometries.
Therefore, employing a $\mathbf{P}_2$ finite element discretization,
we get eight different stencils 
per tetrahedron, including the vertex-centered stencil, see \cref{fig:p2-stencils}.
This uniform structure is also present on the interface-primitives.
Therefore, a matrix-free implementation of classical point-smoothers such as the Gauss-Seidel method can be realized for 
all unknowns.

While all groups of unknowns (including vertex unknowns) can be organized in a tetrahedral
layout, the number of unknowns in one tetrahedron differs per group.
Since the number of vertices on each outer edge of the refined tetrahedron is equal,
we define
\begin{align}\label{eq:num-unknowns-tet}
    N_\text{tet}(v) := ((v+2)(v+1)v)/6
\end{align}
as the number of vertices in a refined tetrahedron with $v$ vertices on the outer edge.
\Cref{tab:unknown-count} lists the number of unknowns for each sub-group.
\begin{table}[ht]
    \centering
    \footnotesize
    \begin{tabular}{r|rr}
        sub-group    & inner unknowns & including boundary \\
        \hline
        vertices    & $N_\text{tet}(2^\ell - 3)$ & $N_\text{tet}(2^\ell + 1)$ \\
        xyz-edges   & $N_\text{tet}(2^\ell - 1)$ & $N_\text{tet}(2^\ell - 1)$ \\
        other edges & $N_\text{tet}(2^\ell - 2)$ & $N_\text{tet}(2^\ell)$     \\
        \hline
    \end{tabular}
    \caption{Number of unknowns in a tetrahedron, with and without boundary.}
    \label{tab:unknown-count}
\end{table}
Note, that the number of unknowns of a $\mathbf{P}_2$ finite-element
discretization on level $\ell$ is equal to the numbers of unknowns of a $\mathbf{P}_1$ finite-element
discretization on level $\ell + 1$.

To estimate the number of floating-point operations that are required to perform matrix-vector
multiplications with the different blocks of the discrete Stokes operator in \cref{eq:stokes-discrete}
we count the stencil entries, \ie the number of non-zero entries in the respective row of the system matrix.
We only consider the stencils in the refined macro-cells, since in the limit, most unknowns are located there.

For different discretizations, the block-matrices of the Stokes operator map from and to
different function spaces.
Therefore we need to consider subsets of the stencils that are
depicted in \cref{fig:p2-stencils} for different operators.
In \cref{tab:stencil-count} we list
the number of stencil entries for the 8 different stencils.
\begin{table}
    \centering
    \footnotesize
    \begin{tabular}{rr|rrrrrrrr}
        && \multicolumn{7}{c}{to (stencil center)} \\
                                    && vertex  & x-e & y-e & z-e & xy-e & xz-e & yz-e & xyz-e \\
        \hline
        \multirow{2}{*}{from} & vertex    & 15 & 8 & 6 & 8 & 8 & 6 & 8 & 6 \\
                              & all  & 65 & 27 & 19 & 27 & 27 & 19 & 27 & 19 \\
        \hline
    \end{tabular}
    \caption{Number of stencil entries for the 8 different stencils as shown in \cref{fig:p2-stencils}.
    The first row only includes the stencil entries that are located on vertex unknowns.}
    \label{tab:stencil-count}
\end{table}

\section{Matrix-free geometric multigrid}\label{sec:multigrid}

To solve the saddle point problem \cref{eq:stokes-discrete}
we employ a monolithic geometric multigrid method. 
By monolithic, we express that the method acts directly on the full system, and not on individual
blocks of the system matrix. 
Traditionally, the Stokes system is often solved 
using Krylov subspace methods such as the Schur-complement
conjugate gradient method or preconditioned MINRES solvers \cite{elman2014finite},
which might in turn employ multigrid on the positive definite submatrix $\mathbf{A}$ of $\mathcal{A}$
in \cref{eq:stokes-discrete}.
Both alternatives  
have been compared to a monolithic multigrid implementation on \gls*{hhg} in
\cite{gmeiner2016quantitative},
where it was shown that 
the monolithic multigrid solver outperforms
both Krylov methods in terms of time-to-solution and
memory consumption.
Therefore we will 
here focus solely on the monolithic multigrid approach,
extending the results in \cite{gmeiner2016quantitative,gmeiner2015towards} 
to the stable mixed $\mathbf{P}_2 - \mathbf{P}_1$ discretization.

\subsection{Full multigrid}

We focus on the \gls*{fmg} scheme
to achieve an asymptotically optimal 
complexity \cite{brandt2011multigrid,hackbusch1994iterative,trottenberg2000multigrid}. 
The algorithmic structure of the \gls*{fmg} algorithm is described in \cref{alg:fmg}.
The number of smoothing iterations performed by the embedded \emph{variable} v-cycle 
increases on the coarser grids. As described in \cite{drzisga2018analysis}, we observe that
this is necessary to achieve satisfactory convergence.

\begin{algorithm}
    \setstretch{1.2}
    \footnotesize
    \algloopdefx{Repeat}[1]{\textbf{repeat} #1 \textbf{times}}
    \begin{algorithmic}[1]
        \Procedure{FMG}{$\nu_\text{pre}, \nu_\text{post}, \nu_\text{inc}, \kappa, \bar{A}, \bar{\xi}$}
            \State solve $\mathcal{A}_0 \underline{\mathbf{x}}_0 = \underline{\mathbf{b}}_0$
            \For{$\ell = 1$ \textbf{to} $L$}
                \State $\underline{\mathbf{x}}_\ell = I_{\ell-1}^{\ell} \underline{\mathbf{x}}_{\ell-1}$
                    \Comment{FMG interpolation}
                \Repeat{$\kappa$}
                    \State $\underline{\mathbf{x}}_\ell \gets$ 
                            \Call{VAR-V-CYCLE}{$\ell, \ell, \nu_\text{pre}, \nu_\text{post}, \nu_\text{inc},
                            \bar{A}, \bar{\xi}, \underline{\mathbf{x}}_\ell, \underline{\mathbf{b}}_\ell$}
            \EndFor
            \State \textbf{return} $\underline{\mathbf{x}}_L$
        \EndProcedure
        \Statex
        \Procedure{VAR-V-CYCLE}{$\ell_\text{fine}, \ell, \nu_\text{pre}, \nu_\text{post}, \nu_\text{inc}, \bar{A}, \bar{\xi},
                \underline{\mathbf{x}}_\ell, \underline{\mathbf{b}}_\ell$}
            \If{$\ell = 0$}
                \State solve $\mathcal{A}_0 \underline{\mathbf{x}}_0 = \underline{\mathbf{b}}_{0}$
                        \Comment{coarse grid solve}
            \Else
                \State $\nu_\text{pre} + (\ell_\text{fine} - \ell) \nu_\text{inc}$ iterations of \cref{eq:uzawa-iteration} on 
                        $\underline{\mathbf{x}}_\ell$ with $(\hat{A}, \xi) = (\bar{A}, \bar{\xi})$
                        \Comment{relaxation}
                \State $\underline{\mathbf{r}}_{\ell-1} = I_\ell^{\ell-1}\left( \underline{\mathbf{b}}_\ell - 
                        \mathcal{A}_\ell \underline{\mathbf{x}}_\ell \right)$
                        \Comment{defect restriction}

                    \State $\hat{\underline{\mathbf{x}}}_{\ell-1} \gets$
                        \Call{VAR-V-CYCLE}{$\ell_\text{fine}, \ell - 1, \nu_\text{pre}, \nu_\text{post}, \nu_\text{inc}, 
                            \bar{A}, \bar{\xi}, 0, \underline{\mathbf{r}}_{\ell-1}$}

                \State $\underline{\mathbf{x}}_{\ell} \gets
                        \underline{\mathbf{x}}_{\ell} + 
                        I_{\ell-1}^{\ell} \hat{\underline{\mathbf{x}}}_{\ell-1}$
                        \Comment{prolongation and correction}

                \State $\nu_\text{post} + (\ell_\text{fine} - \ell) \nu_\text{inc}$ iterations of \cref{eq:uzawa-iteration} on
                        $\underline{\mathbf{x}}_\ell$ with $(\hat{A}, \xi) = (\bar{A}, \bar{\xi})$
                        \Comment{relaxation}
            \EndIf
            \State{ \textbf{return} $\underline{\mathbf{x}}_\ell$}
        \EndProcedure
    \end{algorithmic}
    \caption{\footnotesize \gls*{fmg} algorithm to solve \cref{eq:stokes-discrete} on level $L$. For readability, we set 
    $\underline{\mathbf{x}}_\ell := (\underline{\mathbf{u}}_\ell, \underline{\mathbf{p}}_\ell)^\top$, and
    $\underline{\mathbf{b}}_\ell := (\underline{\mathbf{f}}_\ell, \underline{\mathbf{g}}_\ell)^\top$.}
    \label{alg:fmg}
\end{algorithm}

\subsection{Relaxation}\label{sec:relaxation}

Because of its saddle-point structure, the system of \cref{eq:stokes-discrete} is not positive definite, 
and classical smoothers for scalar elliptic problems cannot be employed directly.
Numerous different smoothers have been developed that can treat the saddle-point problem in
a multigrid framework such as inexact Uzawa-type smoothers 
\cite{gaspar2014simple,drzisga2018analysis}, Braess-Sarazin smoothers \cite{braess1997efficient}
and Vanka-type block-smoothers \cite{vanka1986block}.
Often, these methods use preconditioners on the individual block-matrices that require their diagonals.
Chebyshev smoothers can be employed as an efficient alternative \cite{adams2003parallel,may2015scalable};
they do not require the matrix diagonal and can be implemented in tensor-product approaches
using matrix-vector products 
that can be realized efficiently on modern parallel architectures 
\cite{fehn2018efficiency}.
In the \gls*{hhg} framework, however, we represent the discrete operators by means of stencils that are either
stored block-wise if the operator is (block-wise) constant, or assembled on-the-fly \cite{bauer2018,bauer2017two}.
Therefore the diagonal entries of the system matrix are available for every row of the matrix, 
and smoothers based on matrix-splitting can be realized.

In this work, we employ an \emph{inexact Uzawa}-type smoother \cite{maitre1985fast,gaspar2014simple,drzisga2018analysis} 
directly to \cref{eq:stokes-discrete}. 
This algorithm has shown excellent performance 
as a smoother
for extreme-scale multigrid
methods on \gls*{hhg} in 
\cite{bauer2019large,bauer2018,gmeiner2016quantitative,gmeiner2015performance}.
It is defined as the iteration (dropping the level index $\ell$ in this section)
\begin{align}\label{eq:uzawa-iteration}
\begin{bmatrix}
\underline{\mathbf{u}}_{k+1} \\
\underline{\mathbf{p}}_{k+1}
\end{bmatrix}
=
\begin{bmatrix}
\underline{\mathbf{u}}_k \\
\underline{\mathbf{p}}_k
\end{bmatrix} + 
\mathcal{P}^{-1} \left(
\begin{bmatrix}
\underline{\mathbf{f}} \\
\underline{\mathbf{g}}
\end{bmatrix} - \mathcal{A}
\begin{bmatrix}
\underline{\mathbf{u}}_k \\
\underline{\mathbf{p}}_k
\end{bmatrix} \right), \quad
\mathcal{P} =
\begin{bmatrix}
\hat{\mathbf{A}} & \mathbf{0} \\
\mathbf{B} & - \hat{S}
\end{bmatrix},
\end{align}
where $\mathcal{A}$ is the discrete Stokes system matrix from \cref{eq:stokes-discrete},
and $\hat{\mathbf{A}}$ and $\hat{S}$ are approximations of $\mathbf{A}$ and the Schur complement
$S = \mathbf{B}\mathbf{A}^{-1}\mathbf{B}^\top + C$, respectively.
One iteration of \cref{eq:uzawa-iteration} is then split into two steps:
\begin{align}
\underline{\mathbf{u}}_{k+(j+1)/\xi} &= \underline{\mathbf{u}}_{k+j/\xi} + \hat{\mathbf{A}}^{-1} 
\left( \underline{\mathbf{f}} - \mathbf{A} \underline{\mathbf{u}}_{k+j/\xi} - \mathbf{B}^\top \underline{\mathbf{p}}_k \right),
\label{eq:uzawa-velocity}\\
\underline{\mathbf{p}}_{k+1} &= \underline{\mathbf{p}}_k - \hat{S}^{-1} 
\left( \underline{\mathbf{g}} - \mathbf{B} \underline{\mathbf{u}}_{k+1} + C\mathbf{p}_k \right), 
\label{eq:uzawa-pressure}
\end{align}
with $\xi \in \mathbb{N}^+$, and $j = 0, \dots, \xi-1$.

In a multigrid context, we are interested in the smoothing property of the 
iteration, and choose $\hat{\mathbf{A}}$ and $\hat{S}$ as follows: 
$\mathbf{A}^{-1}$ is approximated by $\xi \geq 1$ Gauss-Seidel iterations,
\ie relaxation on
$\mathbf{A}\underline{\mathbf{u}} = \underline{\mathbf{f}} - \mathbf{B}^\top \underline{\mathbf{p}}$.
We consider different versions for the smoother $\hat{\mathbf{A}}^{-1}$ in \cref{eq:uzawa-velocity}:
a standard Gauss-Seidel relaxation denoted by $\hat{\mathbf{A}} := \hat{\mathbf{A}}_f$, and
a symmetric iteration, \ie one forward and one backward Gauss-Seidel iteration, denoted by 
$\hat{\mathbf{A}} := \hat{\mathbf{A}}_s$.

Regarding the pressure update \cref{eq:uzawa-pressure}, we set $\hat{S} := \omega^{-1} \diag(-C)$ 
where $C$ is the stabilization matrix defined in \cref{eq:pspg} \cite{drzisga2018analysis}.
The relaxation parameter $\omega^{-1}$ is set to an estimate of the maximum absolute eigenvalue of the
eigenvalue problem
\begin{align}\label{eq:omega-eigenvalue-problem}
    M_L^{-1} (C + \mathbf{B} \hat{\mathbf{A}}^{-1}_s \mathbf{B}^\top) \underline{\mathbf{v}} = \lambda \underline{\mathbf{v}}
\end{align}
where $M_L$ is the lumped mass-matrix resulting from a linear finite element
discretization \cite{drzisga2018analysis}. 
For the analysis of the smoothing property of the inexact Uzawa iteration, we refer to \cite{gaspar2014simple,drzisga2018analysis}.

\subsection{Grid transfer}\label{sec:grid-transfer}
The interpolation $I_\ell^{\ell+1}$ from level $\ell$ to $\ell+1$ is chosen according to the order of the
finite-element discretization of the prolongated function. For functions discretized with $\mathbf{P}_2$
finite elements, we employ quadratic interpolation, linear interpolation otherwise. The velocity and pressure
functions are interpolated independently. The restriction is defined as $I^\ell_{\ell+1} := (I_\ell^{\ell+1})^\top$.

\section{Textbook multigrid efficiency}\label{sec:tme}

As introduced in \cref{sec:introduction}, our \gls*{fmg} method achieves \gls*{tme}, if it solves the system of interest, 
in this case \cref{eq:stokes-discrete}, with
\begin{align}
    \mathfrak{W}(\text{FMG})\ / \ \mathfrak{W}(\mathcal{A}) < 10.
\end{align}
In this section, we quantify the computational cost of the proposed \gls*{fmg} iteration in \cref{alg:fmg} for both,
the equal-order and Taylor-Hood finite-element discretizations, to find configurations that achieve \gls*{tme}.

\subsection{Operator application}

We derive a model for the computational cost 
of our FMG solver, starting with an approximation of the computational cost 
of the application of a linear operator.
Therefore, we count the number of involved arithmetic operations.
In the following, we compute the cost of operator applications only in the interior of
a tetrahedron on refinement level $\ell$. 
The number of unknowns in the interior of a tetrahedron is calculated using \cref{eq:num-unknowns-tet}
and \cref{tab:unknown-count}.

For the application of a stencil with $n$ entries, we account $n$ multiplications and $n-1$ additions. 
We approximate this with $2n$ operations for each stencil application.
Where necessary, the number of operations is then multiplied by three, accounting for the three velocity components.

The computational costs for the application of the block operators of \cref{eq:stokes-discrete} 
for the $\mathbf{P}_2 - \mathbf{P}_1$ discretization on level $\ell$ are (\cf \cref{tab:unknown-count,tab:stencil-count})
\begin{align}
    \mathfrak{W}(\mathbf{B}_\ell^{\mathbf{P}_2 - \mathbf{P}_1}) &= \overbrace{3}^{\text{\footnotesize vel. components}} \cdot \ (\underbrace{2}_{\text{\footnotesize add + mul}} \cdot 
    \overbrace{65}^{\text{\footnotesize stencil size}}) \cdot \underbrace{N_\text{tet}(2^{\ell} - 3)}_{\text{\footnotesize inner unknowns}} \label{eq:cost-block-p2p1-b} \\
    \mathfrak{W}(\mathbf{A}_\ell^{\mathbf{P}_2 - \mathbf{P}_1}) &= 3 \cdot
    \left( (2 \cdot 65) \cdot N_\text{tet}(2^{\ell} - 3) + (2\cdot146) \cdot N_\text{tet}(2^{\ell} - 2) \right. \label{eq:cost-block-p2p1-a} \label{eq:cost-block-p2p1-a} \label{eq:cost-block-p2p1-a}\\
    & \phantom{= 3 \cdot (} \left. + (2\cdot19) \cdot N_\text{tet}(2^{\ell} - 1) \right)  \\
    \mathfrak{W}(C_\ell^{\mathbf{P}_2 - \mathbf{P}_1}) &= 0.  \label{eq:cost-block-p2p1-c}
\end{align}
The $\mathbf{P}_1 - \mathbf{P}_1$ finite-element discretization leads to a 15-point stencil in the
interior of a refined tetrahedron \cite{gmeiner2015performance}.
It corresponds to the vertex-centered stencil in \cref{fig:p2-stencils-pt-1-vertex}
without the entries on the edges.
The computational costs for the individual blocks are therefore
\begin{align}
    \mathfrak{W}(\mathbf{A}_\ell^{\mathbf{P}_1 - \mathbf{P}_1}) &= \mathfrak{W}(\mathbf{B}_\ell^{\mathbf{P}_1 - \mathbf{P}_1}) = 3 \cdot  (2\cdot15) \cdot N_\text{tet}(2^{\ell} - 3) \label{eq:cost-block-p1p1-ab} \\
    \mathfrak{W}(C_\ell^{\mathbf{P}_1 - \mathbf{P}_1}) &= (2 \cdot 15) \cdot N_\text{tet}(2^{\ell}-3) \label{eq:cost-block-p1p1-c}.
\end{align}
While $\mathfrak{W}(\mathbf{B}_\ell) \neq \mathfrak{W}(\mathbf{B}_\ell^\top)$ because of boundary effects, 
they asymptotically incur the same cost, \ie we approximate 
$\mathfrak{W}(\mathbf{B}_\ell) \approx \mathfrak{W}(\mathbf{B}_\ell^\top)$.

The cost of an application of a block matrix is approximated with the sum of the work
for the application of the individual blocks. For $\mathcal{A}_\ell$ of \cref{eq:stokes-discrete}
we set
\begin{align}
\mathfrak{W}(\mathcal{A}_\ell) \approx \mathfrak{W}(\mathbf{A}_\ell) + \mathfrak{W}(\mathbf{B}^\top_\ell) + \mathfrak{W}(\mathbf{B}_\ell) + \mathfrak{W}(C_\ell).
\end{align}

To set these numbers into perspective, 
we compare the cost of the application of the Stokes operator 
discretized with $\mathbf{P}_2 - \mathbf{P}_1$ finite elements on level $\ell$ 
with the cost of the operator application
for the \gls*{pspg} stabilized $\mathbf{P}_1 - \mathbf{P}_1$ discretization on level $\ell + 1$.
As stated in \cref{sec:discretization}, the number of unknowns for a $\mathbf{P}_2$
finite-element discretization on level $\ell$ equals the number of unknowns for a $\mathbf{P}_1$
discretization on level $\ell+1$. However, in both formulations for the Stokes problem,
the pressure is discretized linearly. The asymptotic ratio of unknowns 
($\mathbf{P}_2 - \mathbf{P}_1 / \mathbf{P}_1 - \mathbf{P}_1$) for a level ratio $\ell/(\ell+1)$
is (\cf \cref{tab:unknown-count})
\begin{align}
    \lim \limits_{\ell \to \infty} 
    \frac{3 \cdot (N_\text{tet}(2^\ell - 3) + 6 \cdot N_\text{tet}(2^\ell - 2) + N_\text{tet}(2^\ell - 1)) + N_\text{tet}(2^\ell - 3)}
    {4 \cdot N_\text{tet}(2^{\ell+1} - 3)} = \frac{25}{32}.
\end{align}

The asymptotic, work ratios of the individual operator blocks of \cref{eq:stokes-discrete} and the Stokes operators are
\begin{align}\label{eq:wu-comparison-p2p1-p1p1}
    \lim \limits_{\ell \to \infty} 
    \frac{\mathfrak{W}(\mathbf{A}_{\ell}^{\mathbf{P}_2 - \mathbf{P}_1})}
    {\mathfrak{W}(\mathbf{A}_{\ell+1}^{\mathbf{P}_1 - \mathbf{P}_1})} = \frac{23}{12}, \quad
    \lim \limits_{\ell \to \infty} 
    \frac{\mathfrak{W}(\mathbf{B}_{\ell}^{\mathbf{P}_2 - \mathbf{P}_1})}
    {\mathfrak{W}(\mathbf{B}_{\ell+1}^{\mathbf{P}_1 - \mathbf{P}_1})} = \frac{13}{24}, \quad
    \lim \limits_{\ell \to \infty} 
    \frac{\mathfrak{W}(\mathcal{A}_{\ell}^{\mathbf{P}_2 - \mathbf{P}_1})}
    {\mathfrak{W}(\mathcal{A}_{\ell+1}^{\mathbf{P}_1 - \mathbf{P}_1})} = \frac{9}{10}.
\end{align}
Concluding, the work for the application of the two operators 
$\mathcal{A}_{\ell}^{\mathbf{P}_2 - \mathbf{P}_1}$ and $\mathcal{A}_{\ell+1}^{\mathbf{P}_1 - \mathbf{P}_1}$ is comparable.

\subsection{Multigrid}

It follows the estimation of the computational work of the \gls*{fmg} algorithm as listed in \cref{alg:fmg}. 

\paragraph{Relaxation} 

First, we only regard the work of the relaxation \cref{eq:uzawa-iteration},
which we denote as $\mathfrak{W}(\mathcal{P}_\ell^{-1})$.
The velocity update \cref{eq:uzawa-velocity} is decomposed into the application of the discrete gradient $\mathbf{B}_\ell^\top$ 
and the relaxation with $\hat{\mathbf{A}}_\ell$. The pressure update \cref{eq:uzawa-pressure} is split into the 
application of the discrete divergence $\mathbf{B}_\ell$ and multiplication with $\hat{S}_\ell^{-1}$. 
We neglect the application of $\hat{S}_\ell^{-1}$ in our calculations, since it corresponds to a diagonal scaling.

The computational work for a single (forward or backward) Gauss-Seidel iteration is approximated
with the computational work of the application of the corresponding operator. The inverse of the diagonal 
entries of $\mathbf{A}_\ell$ can be precomputed and stored since we are considering constant
coefficients, therefore no additional division has to be counted. We set (dropping the level index $\ell$
for readability)
\begin{align}
    \mathfrak{W}(\mathbf{A}) \approx \mathfrak{W}(\hat{\mathbf{A}}_f^{-1}) 
    = \frac{1}{2} \mathfrak{W}(\hat{\mathbf{A}}_{s}^{-1}).
\end{align}
The cost for one iteration of the Uzawa relaxation \cref{eq:uzawa-iteration} is approximated by
\begin{align}
\mathfrak{W}(\mathcal{P}^{-1}_\ell) \approx \mathfrak{W}(\hat{\mathbf{A}}_\ell^{-1}) + \mathfrak{W}(\mathbf{B}_\ell^\top) + \mathfrak{W}(\mathbf{B}_\ell) + \mathfrak{W}(C_\ell).
\end{align}

\paragraph{Variable v-cycle}

We now turn our attention to the cost analysis of the variable v-cycle as defined in \cref{alg:fmg}. 
The goal is to estimate the cost of one iteration by the work that is performed on the finest grid.
The computational work of a single variable v-cycle iteration with finest level $L$ is denoted 
by $\mathfrak{W}(V^L)$.
The work that is performed only on level $\ell$, is denoted by $\mathfrak{W}(V^L_\ell)$, \ie
$\mathfrak{W}(V^L) = \sum_{\ell = 0}^{L}\mathfrak{W}(V^L_\ell)$.

We neglect the computational work that is required to compute the 
exact solution on the coarsest level $\ell = 0$
since it is, in theory, given a sufficiently coarse grid, 
not a performance bottleneck. Also, the grid transfer is neglected in our model,
since in practice, the execution time is strongly dominated by the remaining components.
For the residual calculation, we account the cost of one operator evaluation $\mathfrak{W}(\mathcal{A}_\ell)$.
Overall the cost of the variable v-cycle on a single level $\ell \in \{ 0, \dots, L\}$ is
\begin{align}
    \mathfrak{W}(V_\ell^L) &= (\nu_\text{pre} + \nu_\text{post} + 2(L - \ell) \nu_\text{inc}) \ 
    \mathfrak{W}(\mathcal{P}_\ell^{-1}) + \mathfrak{W}(\mathcal{A}_\ell).
\end{align}
Recursively, we approximate (asymptotically, the number of
grid points increases by a factor of $8$ per level)
\begin{align}
    \mathfrak{W}(V_{L-q}^L) \approx \frac{1}{8^q}\left(\mathfrak{W}(V_L^L) + 2q\nu_\text{inc}\mathfrak{W}(\mathcal{P}_L^{-1})\right), 
    \quad q \in \{0, \dots, L\}
\end{align}
and provide an upper bound $\mathfrak{W}^*(V^L)$ to the cost of a variable v-cycle on level $L$ by
\begin{align}\label{eq:bound-var-v-cycle}
    \mathfrak{W}(V^L) &=\sum_{\ell = 0}^{L} \mathfrak{W}(V_\ell^L)
    \lesssim \sum_{q = 0}^{\infty} \frac{1}{8^q} \left( \mathfrak{W}(V_L^L) + 2q\nu_\text{inc} \mathfrak{W}(\mathcal{P}_L^{-1}) \right) \\
    & = \frac{8}{7} \mathfrak{W}(V_L^L) + \frac{16}{49} \nu_\text{inc} \mathfrak{W}(\mathcal{P}_L^{-1}) =: \mathfrak{W}^*(V^L).
\end{align}

\paragraph{Full multigrid}

Using \cref{eq:bound-var-v-cycle}, and $\mathfrak{W}(V_\ell^\ell) \approx 8\mathfrak{W}(V_{\ell-1}^{\ell-1})$, 
an upper bound to the cost of the \gls*{fmg} iteration with finest level $L$ is given by
\begin{align}\label{eq:cost-var-fmg}
    \mathfrak{W}(\text{FMG}^L) &= \sum \limits_{\ell = 0}^{L} \kappa \mathfrak{W}(V^\ell)
    \lesssim \kappa \sum \limits_{q = 0}^{\infty} \frac{1}{8^q} \left(\frac{8}{7} \mathfrak{W}(V_L^L) + \frac{16}{49} \nu_\text{inc} \mathfrak{W}(\mathcal{P}_L^{-1}) \right) \\
    & = \frac{8\kappa}{7} \mathfrak{W}^*(V^L).
\end{align}

\paragraph{Normalization to work units}

Our goal is to express the cost of the \gls*{fmg} iteration in \gls*{wu} as defined in \cref{eq:work-unit}.
The cost in \gls*{wu} of an operator or iterative scheme $D$ normalized by the cost of an operator $H$ 
is in the following denoted by
\begin{align}
    \mathfrak{W}_{H}(D) := \mathfrak{W}(D) \ / \ \mathfrak{W}(H).
\end{align}
Of particular interest, is normalization with regards to the work of 
$\mathcal{A}^{\mathbf{P}_2 - \mathbf{P}_1}$ and $\mathcal{A}^{\mathbf{P}_1 - \mathbf{P}_1}$.
Using \cref{eq:cost-block-p2p1-a,eq:cost-block-p2p1-b,eq:cost-block-p2p1-c}, the normalized cost of the 
individual blocks and smoothers are
\begin{align}\label{eq:block-cost-normalized-p2p1}
    \mathfrak{W}_{\mathcal{A}^{\mathbf{P}_2 - \mathbf{P}_1}}(\mathbf{A}^{\mathbf{P}_2 - \mathbf{P}_1}) 
    \xrightarrow[\ell \to \infty]{} \frac{46}{72}, \quad
    \mathfrak{W}_{\mathcal{A}^{\mathbf{P}_2 - \mathbf{P}_1}}(\mathbf{B}^{\mathbf{P}_2 - \mathbf{P}_1}) 
    \xrightarrow[\ell \to \infty]{} \frac{13}{72}.
\end{align}
for the $\mathbf{P}_2 - \mathbf{P}_1$ discretization, and 
\begin{align}\label{eq:block-cost-normalized-p1p1}
    \mathfrak{W}_{\mathcal{A}^{\mathbf{P}_1 - \mathbf{P}_1}}(\mathbf{A}^{\mathbf{P}_1 - \mathbf{P}_1}) 
    \xrightarrow[\ell \to \infty]{} \frac{3}{10}, \quad
    \mathfrak{W}_{\mathcal{A}^{\mathbf{P}_1 - \mathbf{P}_1}}(C^{\mathbf{P}_1 - \mathbf{P}_1}) 
    \xrightarrow[\ell \to \infty]{} \frac{1}{10}
\end{align}
with $\mathfrak{W}_{\mathcal{A}^{\mathbf{P}_1 - \mathbf{P}_1}}(\mathbf{A}^{\mathbf{P}_1 - \mathbf{P}_1}) 
=\mathfrak{W}_{\mathcal{A}^{\mathbf{P}_1 - \mathbf{P}_1}}(\mathbf{B}^{\mathbf{P}_1 - \mathbf{P}_1})$ 
in the \gls*{pspg} stabilized case (\cf \cref{eq:cost-block-p1p1-ab,eq:cost-block-p1p1-c}).

For the $\mathbf{P}_2 - \mathbf{P}_1$ finite element discretization, 
the cost of an Uzawa relaxation $\mathfrak{W}_{\mathbf{P}_2 - \mathbf{P}_1}(\mathcal{P}^{-1})$
using a forward Gauss-Seidel relaxation ($\hat{\mathbf{A}} = \hat{\mathbf{A}}_f,\ \xi = 1$)
and a symmetric version ($\hat{\mathbf{A}} = \hat{\mathbf{A}}_s,\ \xi = 1$) therefore account for 
$1$ WU and $\frac{118}{72}$ WU respectively.

Exemplarily, the cost of the \gls*{fmg} iteration for the $\mathbf{P}_2 - \mathbf{P}_1$ discretization 
of the Stokes problem parameterized as $\text{FMG}(1, 1, 2, 1, \hat{\mathbf{A}}_s, 1)$ (\cf \cref{alg:fmg})
can be calculated as
\begin{align}
    \mathfrak{W}_{\mathbf{P}_2 - \mathbf{P}_1}(\text{FMG}(1, 1, 2, 1, \hat{\mathbf{A}}_s, 1)) = 
    \frac{8}{7} (\frac{8}{7} ( (1+1+0) \frac{118}{72} + 1 ) + \frac{16}{49} \cdot 2 \cdot \frac{118}{72} ) \approx 6.8
\end{align}
which would, if the system were solved sufficiently, indicate \gls*{tme} according to \cref{eq:tme}.
In the following section, we experiment with different configurations to assess the quality
of our solver with respect to \gls*{tme}.

\section{Performance analysis of the multigrid solver}\label{sec:performance}

In this section, we address the numerical efficiency and computational performance of the 
inexact-Uzawa multigrid solver. 
Two benchmarks are considered to assess the efficiency of our solver with
regard to \gls*{tme}. Finally we present a roofline analysis of the employed
relaxation kernels and demonstrate the extreme scalability of the implementation.

All benchmarks in this section are performed and reproducible with the open source framework 
\hyteg \footnote{\url{https://i10git.cs.fau.de/hyteg/hyteg}, Git-SHA: a8e393e8} \cite{kohl2019hyteg}, 
which implements all data structures and solvers described in this article.
For the coarse grid solvers, we utilize the \gls*{petsc}\footnote{\url{https://www.mcs.anl.gov/petsc/}} 
\cite{Petsc1997}.

\subsection{Numerical efficiency}\label{sec:numerical-efficiency}

The convergence rate of an iterative solver alone does not permit any
statement about its numerical efficiency, as long as the involved computational work
is not considered. In this section, both metrics, \ie convergence
and computational work are combined to assess, and optimize numerical efficiency
of the proposed multigrid solver, and, at best, reach \gls*{tme}.

Let $x$ define a continuous, exact solution for the underlying \gls*{pde}, 
$\underline{\mathbf{x}}^*_{\ell,j}$ 
the evaluation of $x$ at grid node $j$ on level $\ell$, 
and $I_\ell^{\ell+1}$ as in \cref{sec:grid-transfer}.
The discrete $L^2$-error $\norm{\mathbf{e}(\tilde{\underline{\mathbf{x}}}_\ell)}_2$ of a computed solution $\tilde{\underline{\mathbf{x}}}_\ell$ 
is in the following defined as
\begin{align}\label{eq:error}
    \norm{\mathbf{e}(\tilde{\underline{\mathbf{x}}}_\ell)}_2 := 
    \norm{\underline{\mathbf{x}}^*_{\ell+1} - I_\ell^{\ell+1} \tilde{\underline{\mathbf{x}}}_\ell}_2, \quad
    \norm{\underline{\mathbf{x}}_\ell}_2 = \sqrt{\underline{\mathbf{x}}_\ell^\top\mathbf{M}_\ell\underline{\mathbf{x}}_\ell},
\end{align}
where $\mathbf{M}_\ell$ is the corresponding finite-element mass-matrix on level $\ell$.
The interpolation to a finer grid is performed to reflect the continuity of the 
finite-element solution in the error.
Let $\underline{\mathbf{x}}_\ell$ a solution of the considered linear system 
$\mathcal{A}\underline{\mathbf{x}}_\ell = \underline{\mathbf{b}}_\ell$ with
$(\underline{\mathbf{r}}_u, \underline{\mathbf{r}}_p)^\top = \underline{\mathbf{b}}_\ell - \mathcal{A}\underline{\mathbf{x}}_\ell$ and
$\norm{\underline{\mathbf{r}}_u}_2 < \epsilon \land \norm{\underline{\mathbf{r}}_p}_2 < \epsilon$.
We define $\gamma(\tilde{\underline{\mathbf{x}}}_\ell)$ as
the ratio of the error of a computed
solution $\tilde{\underline{\mathbf{x}}}_\ell$ to the discretization error,
\begin{align}\label{eq:gamma}
    \gamma(\tilde{\underline{\mathbf{x}}}_\ell) := 
    \norm{\mathbf{e}(\tilde{\underline{\mathbf{x}}}_\ell)}_2 / \norm{\mathbf{e}(\underline{\mathbf{x}}_\ell)}_2.
\end{align}
We set for our experiments $\epsilon = 10^{-12}$.
A well-parameterized \gls*{fmg} solver should reduce the error of the solution, so that
$\gamma(\tilde{\underline{\mathbf{x}}}_\ell)$ is close to $1$ 
\cite{trottenberg2000multigrid,gmeiner2016quantitative}.

Our aim is to achieve
\gls*{tme}, \ie find parameterizations $s$ with 
$\mathfrak{W}(\text{FMG}(s)) < 10$, for which
the \gls*{fmg} iteration \emph{solves the underlying \gls*{pde}}
up to discretization accuracy. Solving the discrete system exactly is
usually much more expensive, however, not relevant, since the discretization error dominates
the approximation.
We loosely consider the \gls*{pde} \cref{eq:stokes-cc} solved, if the computed solution
$(\underline{\mathbf{u}}_\ell, \underline{\mathbf{p}}_\ell)$ of \cref{eq:stokes-discrete} satisfies 
$\gamma(\underline{\mathbf{u}}_\ell),\gamma(\underline{\mathbf{p}}_\ell) \leq 2$. 
Clearly, lower bounds can be achieved, and may be desired, too.
In general, we expect $\mathfrak{W}(\text{FMG}(s))$ to inversely depend on the 
chosen bounds to $\gamma(\underline{\mathbf{u}}_\ell)$ and $\gamma(\underline{\mathbf{p}}_\ell)$.
Finding a \emph{numerically efficient} \gls*{fmg} solver, can be viewed as a multi-objective
optimization problem, in this case with objectives $\mathfrak{W}(\text{FMG}(s))$, 
$\gamma(\underline{\mathbf{u}}_\ell)$, and 
$\gamma(\underline{\mathbf{p}}_\ell)$.

To find efficient parameterizations, we define a parameter search space 
$S = \{(\bar{\nu}_\text{pre},\allowbreak
\bar{\nu}_\text{post},\allowbreak
\bar{\nu}_\text{inc},\allowbreak
\bar{\kappa},\allowbreak
\bar{\mathbf{A}}, \bar{\xi}) : \allowbreak
0 \leq \bar{\nu}_\text{pre}, \bar{\nu}_\text{post}, \bar{\nu}_\text{inc} \leq 3,\ 
1 \leq \bar{\kappa} \leq 2,\ 
(\bar{\mathbf{A}}, \bar{\xi}) \in 
\{(\hat{\mathbf{A}}_f, 1),\allowbreak
(\hat{\mathbf{A}}_f, 2),\allowbreak
(\hat{\mathbf{A}}_f, 3),\allowbreak
(\hat{\mathbf{A}}_f, 4),\allowbreak
(\hat{\mathbf{A}}_s, 1),\allowbreak
(\hat{\mathbf{A}}_s, 2)
\}\}$, containing $768$ tuples.

We employ the \gls*{mumps} \cite{buttari:hal-02528532}, interfaced through \gls*{petsc} \cite{Petsc1997}
as a direct solver for the unstructured coarse grid ($\ell = 0$).

\subsubsection{Cube with analytical solution}\label{sec:benchmark-cube}

As in \cite{gmeiner2015towards}, we consider an analytical solution for 
\cref{eq:stokes-cc} on the cube $\Omega = (0, 1)^3$:
\begin{align}
    \bm{u} &= 
    (-4 \cos(4 x_3), 8 \cos(8 x_1), -2 \cos(2 x_2))^\top, \\
    p &= \sin( 4 x_1 ) \sin( 8 x_2 ) \sin( 2 x_3 ) + \text{const}
\end{align}
with the right-hand side $\mathbf{f}$ that satisfies \cref{eq:stokes-cc}, 
and $\partial\Omega_D = \partial\Omega$. The cube is partitioned into
24 macro-tetrahedra, that are uniformly refined $L$ times, resulting in about $4.3 \times 10^6$ unknowns for the $\mathbf{P}_1-\mathbf{P}_1$
discretization ($L=6$) and $3.4 \times 10^6$ unknowns for the $\mathbf{P}_2-\mathbf{P}_1$ discretization ($L=5$). The relaxation parameter
$\omega^{-1}$ is estimated by $100$ power iterations on \cref{eq:omega-eigenvalue-problem},
resulting in $\omega^{-1} = 0.448872$ for the $\mathbf{P}_2-\mathbf{P}_1$ and 
$\omega^{-1} = 0.570751$ for the $\mathbf{P}_1-\mathbf{P}_1$ discretization.

In \cref{tab:fmg-table}, we list some configurations found by optimizing for minimal computational
work of the resulting \gls*{fmg} iteration, while prescribing upper bounds $\hat{\gamma}_u$ and $\hat{\gamma}_p$ for 
$\gamma(\underline{\mathbf{u}}_L)$ and $\gamma(\underline{\mathbf{p}}_L)$.
Also, we fixed $\kappa = 1$, since frequent visits of the coarse grids generally decrease the performance
in parallel settings, and allowing $\kappa = 2$ does not substantially reduce the minimal work obtained.

The results for both discretizations are satisfactory, as \gls*{tme} is achieved, or almost achieved for all chosen upper bounds
in \cref{tab:fmg-table}.
Notably, for the $\mathbf{P}_2-\mathbf{P}_1$ case, at least $3$ inner Gauss-Seidel iterations in \cref{eq:uzawa-velocity} 
seem to be favorable, and especially increasing the number of post-smoothing steps appears to improve the efficiency of the iteration.
For both discretizations, a looser bound to the pressure error significantly reduces the required work.
If only a very accurate velocity solution is of interest, very efficient parameterizations may be chosen.
We note that no configuration could reduce the pressure 
error so that $\gamma(\underline{\mathbf{p}}_L) \leq 1.4$.
Similar results for the $\mathbf{P}_1-\mathbf{P}_1$ discretization are presented and discussed in 
\cite{gmeiner2015towards,gmeiner2015performance}.

\begin{table}
\footnotesize
\centering
$\mathbf{P}_1-\mathbf{P}_1,\ L=6,\ \min_{s \in \hat{S}}(\mathfrak{W}(\text{FMG}(s))),\ \hat{S} := \{s \in S : \kappa = 1 \land \gamma(\underline{\mathbf{u}}_L) \leq \hat{\gamma}_u \land \gamma(\underline{\mathbf{p}}_L) \leq \hat{\gamma}_p\}$
\begin{tabular}{rrrr}
\hline
 $(\hat{\gamma}_u, \hat{\gamma}_p)$   & $(1.1, 2)$                            & $(1.1, 5)$                            & $(2, 10)$                             \\
\hline
 $s$                                  & $(2, 3, 2, 1, \hat{\mathbf{A}}_s, 1)$ & $(3, 1, 3, 1, \hat{\mathbf{A}}_s, 1)$ & $(1, 0, 2, 1, \hat{\mathbf{A}}_s, 1)$ \\
 $\mathfrak{W}(\text{FMG}(s))$        & 10.77                                 & 9.55                                  & 3.97                                  \\
 $\gamma(\underline{\mathbf{u}}_L)$   & 1.10                                  & 1.10                                  & 1.62                                  \\
 $\gamma(\underline{\mathbf{p}}_L)$   & 1.51                                  & 2.91                                  & 8.52                                  \\
\hline
\end{tabular}\\

\vspace{6pt}
$\mathbf{P}_2-\mathbf{P}_1,\ L=5,\ \min_{s \in \hat{S}}(\mathfrak{W}(\text{FMG}(s))),\ \hat{S} := \{s \in S : \kappa = 1 \land \gamma(\underline{\mathbf{u}}_L) \leq \hat{\gamma}_u \land \gamma(\underline{\mathbf{p}}_L) \leq \hat{\gamma}_p\}$
\begin{tabular}{rrrr}
\hline
 $(\hat{\gamma}_u, \hat{\gamma}_p)$   & $(1.1, 2)$                            & $(1.1, 5)$                            & $(2, 10)$                             \\
\hline
 $s$                                  & $(1, 3, 2, 1, \hat{\mathbf{A}}_f, 3)$ & $(1, 2, 1, 1, \hat{\mathbf{A}}_f, 3)$ & $(0, 2, 1, 1, \hat{\mathbf{A}}_f, 3)$ \\
 $\mathfrak{W}(\text{FMG}(s))$        & 14.91                                 & 11.08                                 & 8.11                                  \\
 $\gamma(\underline{\mathbf{u}}_L)$   & 1.01                                  & 1.02                                  & 1.47                                  \\
 $\gamma(\underline{\mathbf{p}}_L)$   & 1.99                                  & 4.55                                  & 9.81                                  \\
\hline
\end{tabular}
\caption{Results for some parameterizations from the search space $S$, optimized towards minimal work with fixed upper bounds 
         for $\gamma(\underline{\mathbf{u}}_L)$ and $\gamma(\underline{\mathbf{p}}_L)$, and $\kappa = 1$.}
\label{tab:fmg-table}
\end{table}

\begin{figure}
\centering
    \centering
    \scalebox{0.6}{\input{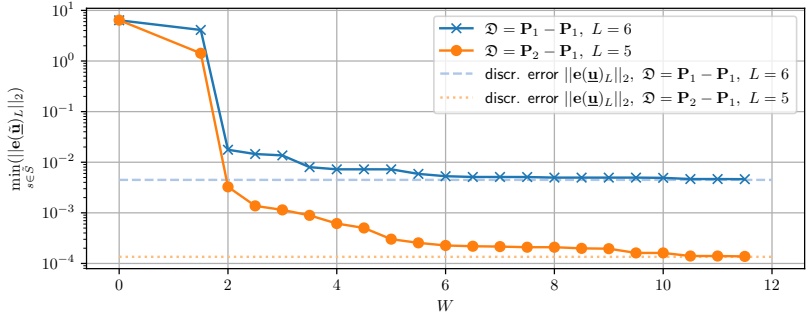}}
    \caption{Minimal achievable velocity error after an \gls*{fmg} iteration with parameterization $s \in S$ 
             that requires a maximum work of $W$, plotted for $W \in \{0\} \cup \{1.5, 2, \dots, 12\}$.
             Precisely, 
             $\min_{s \in \tilde{S}}(\norm{\mathbf{e}(\tilde{\underline{\mathbf{u}}}_L)}_2)$ with 
             $\tilde{S} := \{s \in S : \mathfrak{W}_\mathfrak{D}(\text{FMG}(s)) \leq W\}$, and discretization 
             $\mathfrak{D}$ on level $L$.}
    \label{fig:graph-min-error-for-work}
\end{figure}

\Cref{tab:fmg-table} lacks a direct comparison of the efficiency among the two discretizations.
Again, we are interested in \emph{solving the \gls*{pde}}, and not in the exact solution of the discrete problem. To this end, the
ratios $\gamma(\underline{\mathbf{u}}_\ell)$ and $\gamma(\underline{\mathbf{p}}_\ell)$ fail to express a discretization-invariant, 
quantitative measure for accuracy of the computed solution. 
The ratio \cref{eq:wu-comparison-p2p1-p1p1} suggests, that a similar amount
of work is required to apply either $\mathcal{A}^{\mathbf{P}_2 - \mathbf{P}_1}_\ell$ or $\mathcal{A}^{\mathbf{P}_1 - \mathbf{P}_1}_{\ell+1}$.
We compare therefore the error $\norm{\mathbf{e}(\tilde{\underline{\mathbf{u}}}_L)}_2$ as defined in \cref{eq:error} after applying
the \gls*{fmg} iteration to solve \cref{eq:stokes-discrete} with $\mathcal{A}_L = \mathcal{A}^{\mathbf{P}_2 - \mathbf{P}_1}_L$, $L=5$,
and $\mathcal{A}_{L} = \mathcal{A}^{\mathbf{P}_1 - \mathbf{P}_1}_{L}$ with $L = 6$. In particular, we plot in \cref{fig:graph-min-error-for-work} 
the minimal velocity error $\norm{\mathbf{e}(\tilde{\underline{\mathbf{u}}}_L)}_2$ that can be achieved with an \gls*{fmg} configuration 
that requires a certain maximum amount of work $W$.
For the considered example, the $\mathbf{P}_1-\mathbf{P}_1$ discretization accuracy on level $L = 6$ is reached even for really efficient
configurations of the \gls*{fmg} solver when employing the $\mathbf{P}_2-\mathbf{P}_1$ discretization on level $L=5$. For example, a $10$-fold
reduction of the low-order discretization error is achieved with $\mathfrak{W}_{\mathbf{P}_2-\mathbf{P}_1}(\text{FMG}) \approx 5$.

\subsubsection{Flow through a junction}

\begin{figure}[h]
    \centering
    \includegraphics[width=\textwidth]{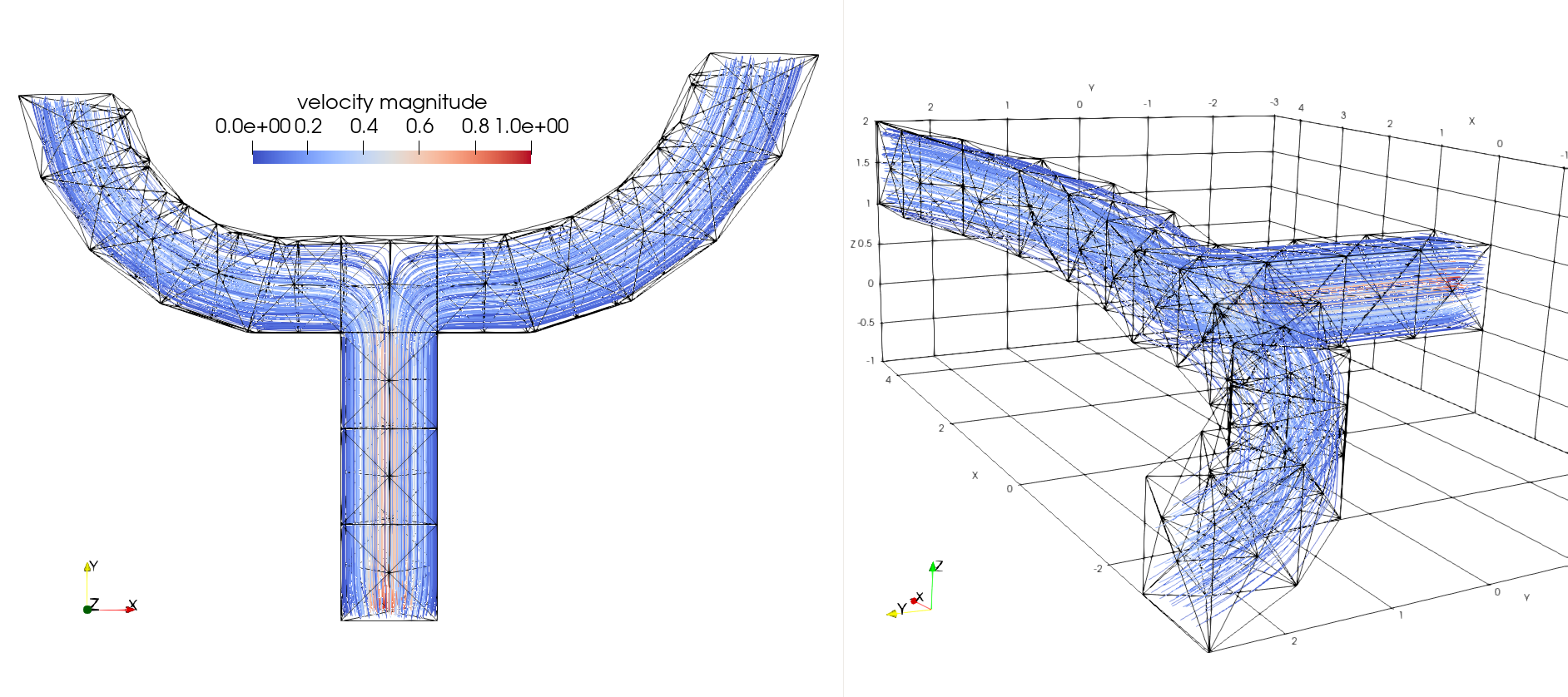}
    \caption{Domain and computed solution of the flow through a y-shaped junction.}
    \label{fig:y-pipe}
\end{figure}

As a second benchmark problem, we consider a y-shaped junction, that is slightly bent in z-direction, with a single, sinusoidal 
inflow and two natural outflow (Neumann) boundaries, \cf \cref{fig:y-pipe}.
For this benchmark the $\mathbf{P}_2-\mathbf{P}_1$ discretization is employed.
The domain consists of $336$ coarse grid tetrahedra, that are
each refined $L = 4$ times, resulting in a system with $5.9 \times 10^6$ unknowns (including boundary). Since no analytical solution is available, we assess
the \gls*{fmg} solver by measuring the relative error $\delta$ of the computed (velocity and pressure) solution $\tilde{\underline{\mathbf{x}}}_\ell$,
to a given solution $\underline{\mathbf{x}}_\ell$ that reduces the velocity and pressure residual so that 
$\norm{\underline{\mathbf{r}}_u}_2 < \epsilon \land \norm{\underline{\mathbf{r}}_p}_2 < \epsilon,\ \epsilon = 10^{-12}$ 
(\cf \cref{sec:numerical-efficiency}), defined as 
\begin{align}\label{eq:relative-error}
\delta(\underline{\mathbf{x}}_\ell) := 
\norm{\tilde{\underline{\mathbf{x}}}_\ell - \underline{\mathbf{x}}_\ell}_2 / \norm{\underline{\mathbf{x}}_\ell}_2.
\end{align}
The estimation of $\omega^{-1}$ with \cref{eq:omega-eigenvalue-problem} does not yield good results, so we set  $\omega^{-1} = 0.2$ empirically.

In \cref{tab:junction-errors}, we list the relative errors for velocity and pressure for the parameterizations found in \cref{tab:fmg-table}.
\begin{table}
\footnotesize
\centering
$\mathbf{P}_2-\mathbf{P}_1,\ L=4$\\
\begin{tabular}{rrrr}
 $s$                                  & $(1, 3, 2, 1, \hat{\mathbf{A}}_f, 3)$ & $(1, 2, 1, 1, \hat{\mathbf{A}}_f, 3)$ & $(0, 2, 1, 1, \hat{\mathbf{A}}_f, 3)$ \\
 $\mathfrak{W}(\text{FMG}(s))$        & 14.91                                 & 11.08                                 & 8.11                                  \\
 \hline
 $\delta(\underline{\mathbf{u}}_L)$   & 0.10\%                                & 0.12\%                                &   0.19\%                              \\
 $\delta(\underline{\mathbf{p}}_L)$   & 5.37\%                                & 6.39\%                                &   9.40\%                              \\
\hline
\end{tabular}
\caption{Relative errors for velocity and pressure as defined in \cref{eq:relative-error} for computed solutions for the junction domain.
         The \gls*{fmg} configurations are chosen as in \cref{tab:fmg-table}.}
\label{tab:junction-errors}
\end{table}
Similarly to the results of the first benchmark problem, the relative velocity error is reduced successfully below $0.2\%$ after only a single \gls*{fmg} iteration. 
The pressure error remains above $5\%$ for all settings, which can be considered insufficient, if the pressure solution is of interest in the 
respective application. Overall, the solver shows a similar qualitative behavior for a different domain.

\subsection{Computational performance}

This last section briefly covers the computational performance on the node-level and 
the parallel scalability. In particular, we are concerned with the implementation of the 
kernels and \gls*{hhg} data structures for the $\mathbf{P}_2-\mathbf{P}_1$ discretization.
The performance and scalability for the linear, equal-order discretization is \eg analyzed in
\cite{gmeiner2015performance,gmeiner2016quantitative}.
We note, that a more concise, in-depth performance analysis on top of the theoretical
results is important, but out of the scope of this article and subject to future work.

\subsubsection{Node-level performance}

Efficient node-level execution is key to fast and scalable numerical code. 
Parallel code based on
slow compute kernels
performs well in scaling experiments, and in fact, 
a poor parallel scalability can be hidden by introducing redundant work
that slows down the kernel performance
\cite{bailey2009misleading,Hager2010introduction}.
Therefore, in our approach
we first strive to achieve the best possible node-level performance
before we
target scalability.

The roofline model \cite{Hager2010introduction,williams2009roofline} provides a simplified,
and optimistic tool to predict the maximal possible performance of a compute kernel implementation.
The corresponding upper bounds are either prescribed by the computational peak performance or the memory bandwidth of the machine.
To analyze the node-level performance of our multigrid implementation, we conduct a roofline analysis 
for the compute kernels of the Gauss-Seidel implementation.

The implementation of the Gauss-Seidel iteration is split into 8 kernels, each of which
updates one group of unknowns as introduced in \cref{sec:edgedofs}. We measure the performance
of each kernel individually, applied to 8 macro-cells in parallel, that are refined 8 times respectively.
The measurements are performed on an Intel\textsuperscript{\textregistered} Xeon\textsuperscript{\textregistered} Gold 5122 CPU,
with 2 sockets, and 8 cores in total. Communication is not included in the measurements. To obtain the performance of the kernels,
and bandwidth of the machine, we use the \likwid\footnote{https://github.com/RRZE-HPC/likwid} \cite{Treibig2010likwid}
tools (version 4.3.4).

\begin{figure}
\centering
    \centering
    \scalebox{0.6}{\input{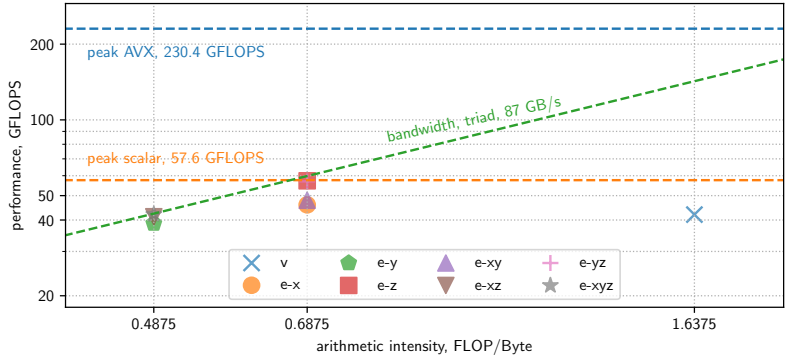}}
    \caption{Roofline plot containing all $8$ kernels employed in the Gauss-Seidel relaxation.}
    \label{fig:roofline}
\end{figure}

In \cref{fig:roofline}, the results of the roofline analysis are plotted, together with the measured bandwidth,
and the theoretical peak, and scalar peak performance of the machine. 
With a fixed clock rate of 3.60 GHz, 8 cores, AVX-256, and 2 fused multiply-adds (FMA) per cycle we get a 
theoretical peak of 230.4 GFLOPS.
All 7 kernels that update the edge unknowns are memory-bound, the implementation reaches the bandwidth limit
in almost all cases. \likwid reports at least $95\%-98\%$ of the performed FLOP originate from AVX instructions,
for the edge-centered kernels.
The kernel applying the vertex-centered stencil is, due to its high arithmetic intensity, not memory bound.
However, it cannot be vectorized because of the lexicographic update pattern and is therefore expected to
be bound by the scalar peak performance; \likwid reports $75\%$ of scalar peak performance.

We emphasize, that the roofline model tends to estimate optimistic upper performance bounds. While it
gives a first impression of the performance, refined approaches like the \gls*{ecm} model provide 
more realistic estimates \cite{stengel2015quantifying}.

\subsubsection{Parallel scalability}

Finally, we demonstrate strong and weak scaling results of the \gls*{fmg} implementation.
\begin{figure}
    \centering
    \centering
    \scalebox{0.6}{\input{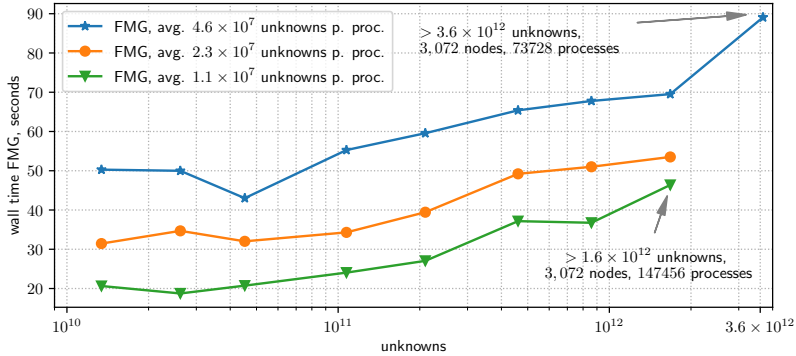}}
    \caption{Weak and strong scaling of the \gls*{fmg} solver with 
    $(\nu_\text{pre}, \nu_\text{post}, \nu_\text{inc}, \kappa, \hat{\mathbf{A}}, \xi) =  (0, 2, 1, 1, \hat{\mathbf{A}}_f, 3)$.}
    \label{fig:scaling}
\end{figure}
The scaling experiments are conducted on the thin nodes of \supermucng, which is ranked 13\textsuperscript{th} in the Top500 as of June 
2020\footnote{\url{https://www.top500.org/}}. Each node is equipped with two
Intel\textsuperscript{\textregistered} Skylake Xeon\textsuperscript{\textregistered}
Platinum 8174 CPU, \ie a total of $48$ cores per node, and $96$GB main memory. In total, the machine operates $6,336$
(thin) nodes; we have access to a maximum of $3,072$ nodes. For the scalability experiment, we run the benchmark as
described in \cref{sec:benchmark-cube}, with finer coarse grids, to balance the number of macro-primitives among the parallel processes, 
and $7$-times refined tetrahedra.

\Cref{fig:scaling} shows the wall time for the \gls*{fmg} solver for the $\mathbf{P}_2-\mathbf{P}_1$ discretization, and
$(\nu_\text{pre}, \nu_\text{post}, \nu_\text{inc}, \kappa, \hat{\mathbf{A}}, \xi) =  (0, 2, 1, 1, \hat{\mathbf{A}}_f, 3)$.
A weak scaling of three configurations is presented: $1.1\times10^{7}$, $2.3\times10^{7}$, and $4.6\times10^{7}$ unknowns per
process on average. For the latter scenario, the number of processes per node is reduced from $48$ to $24$.

We demonstrate scalability to all available $147,456$ processes, and, in the largest scenario, 
solve a Stokes system with more than $3.6 \times 10^{12}$ unknowns in about $90$ seconds. The solution vector
alone requires more than $28$TB of main memory. The monolithic multigrid solver with inexact Uzawa relaxation
is especially suited for large scale computations, as it can be implemented with only one additional temporary vector,
on top of the solution and right-hand side.

We emphasize, that such extreme scalability can only be achieved with matrix-free solvers, and careful choice and 
implementation of the corresponding algorithms and data structures \cite{gmeiner2016quantitative}.

\section{Conclusion}
We demonstrate in this article, that \gls*{tme} can be achieved for different discretizations of the Stokes problem,
while maintaining high computational performance and low memory overhead. The relevant data structures
to realize scalable matrix-free implementations, also for higher-order finite-element discretizations, are introduced. 
Two benchmark problems show, that a satisfactory error reduction of the velocity component can be achieved with computational work
that is of the order of $10$ discrete operator applications. 
The computational efficiency and parallel scalability of the implementation is presented by a roofline analysis, and weak and
strong scaling experiments, demonstrate scalability to up to $147,456$ parallel processes and systems with more than 
$3.6 \times 10^{12}$ unknowns.
This article may serve as a basis for further analysis of
the efficiency of Stokes solvers, in particular for the case of varying coefficients and for coupled, possibly non-linear
applications.
\section*{Acknowledgements}
The authors gratefully acknowledge the Gauss Centre for Supercomputing e.V. 
(\url{https://www.gauss-centre.eu}) for funding this project by providing computing 
time on the GCS Supercomputer SuperMUC-NG at Leibniz Supercomputing Centre (\url{https://www.lrz.de}).

\bibliographystyle{plain}
\bibliography{bibliography}

\end{document}